\documentclass[prl,twocolumn,reprint,showpacs]{revtex4-1}

\usepackage{amsmath,amssymb}

\usepackage{graphicx}

\newcommand{\ad}{a^\dagger}
\newcommand{\an}{a^{}}
\newcommand{\bd}{b^\dagger}
\newcommand{\bn}{b^{}}

\newcommand{\ii}{\mathrm{i}}
\newcommand{\dd}{\mathrm{d}}

\begin{document}

\title{Route to chaos in optomechanics}

\author{L. Bakemeier}
\email{bakemeier@physik.uni-greifswald.de}
\author{A. Alvermann}
\email{alvermann@physik.uni-greifswald.de}
\author{H. Fehske}
\email{fehske@physik.uni-greifswald.de}
\affiliation{Institut f{\"u}r Physik, Ernst-Moritz-Arndt-Universit{\"a}t, 17487 Greifswald, Germany}

\begin{abstract}
We establish the emergence of chaotic motion in optomechanical systems.
Chaos appears at negative detuning for experimentally accessible values of the pump power and other system parameters.
We describe the sequence of period doubling bifurcations that leads to chaos, and state the experimentally observable signatures in the optical spectrum.
In addition to the semi-classical dynamics we analyze the possibility of chaotic motion in the quantum regime.
We find that quantum mechanics protects the optomechanical system against irregular dynamics, such that simple periodic orbits reappear and replace the classically chaotic motion.
In this way observation of the dynamical signatures makes it possible to pin down the crossover from quantum to classical mechanics.
\end{abstract}

\pacs{42.50.Ct,37.10.Vz,05.45.-a,07.10.Cm}

\maketitle

The coupling between light and matter lies at the heart of modern physics.
In recent years the fabrication of optomechanical systems using, e.g.,
microtoroid resonators~\cite{CRYKV05,KRCSV05,RKCV05}, suspended micromirrors~\cite{ACBPH06,G_etal06}, whispering gallery microdisks~\cite{JLRVP06,WCGL09} or microsphere resonators~\cite{MSDDAK07,PW09,TC09}
has opened up new possibilities for fundamental research and technological applications~\cite{KV08,MG09,M13,AKM13}.
Because the light-matter coupling and other system parameters can be adjusted over large scales
optomechanics provides a genuine opportunity to access the classical and quantum dynamics of mesoscopic driven dissipative systems in a variety of different regimes.
Optomechanical systems have been used---or proposed to be used---for the
creation of non-classical light~\cite{B_etal12}, preparation of Schr\"odinger cat states~\cite{Lue_etal13},
generation of light-matter entanglement~\cite{GKPBLS14}, ultra-precision measurements~\cite{ZLFX12,LALM13}, and radiative cooling to the ground state~\cite{T_etal11, C_etal11}.

The basic optomechanical system
consists of a cantilever in a cavity.
The cantilever motion is affected by the radiation pressure of the cavity field, and thus implements light-matter coupling at a truly fundamental level.
The cavity is pumped with an external laser, which drives the system out of equilibrium.
Experiments have successfully demonstrated the optical bistability of the cavity-cantilever dynamics
that leads to self-induced cantilever oscillations~\cite{KRCSV05, MLNOFKM08, ZPASB11}.
With a few exceptions~\cite{CRYKV05, CCV07}, previous studies mainly addressed the regime of simple periodic cantilever motion,
and took the prevalence of regular over irregular dynamics for granted.

In this Letter, we consider the dynamics of the optomechanical system
with a view towards chaotic motion.
We demonstrate the appearance of chaos at negative detuning and explain how to detect it experimentally through characteristic signatures in the optical spectrum.
Chaos emerges already for slightly increased pump power
which makes it accessible with present experimental setups.
We identify the period doubling bifurcations on the way to chaos, and provide the bifurcation diagrams for the first chaotic orbits.

Chaotic dynamics of the optomechanical system appears in the bad-cavity limit and is described by the semi-classical equations of motion.
In the quantum regime we use a Monte Carlo propagation technique~\cite{GP92, D88} to solve the master equation for the density matrix,
which allows us to track the deviations from the classical dynamics systematically.
Surprisingly, chaotic motion can be suppressed in favor of regular oscillatory motion of the cantilever by pushing the system into the quantum regime.
We can relate the reemergence of  periodic cantilever oscillations to the localization of individual quantum trajectories on simple limit cycles that are not accessible in the classical dynamics.

Our theoretical analysis is based on the generic Hamilton operator of optomechanics~\cite{Law95,M13,AKM13}
\begin{equation}
\frac{1}{\hbar} \, H = \left[ -\Delta + g_0(\bn + \bd) \right]\ad\an + \Omega \bd\bn + \alpha_L (\ad + \an) \;.
\label{eq:hamiltonian}
\end{equation}
It describes, e.g., the vibrational mode of a cantilever ($b^{(\dagger)}$, with frequency $\Omega$) under the influence of the radiation pressure ($\propto g_0$) of the cavity photon field $(a^{(\dagger)})$.
To include the effect of the pump laser, with amplitude $\alpha_L$
and detuning $\Delta = \omega_\text{las} - \omega_\text{cav}$
of the laser and cavity frequency,
the Hamilton operator is written in a reference frame rotating at the laser frequency.
To account for radiative cavity losses ($\propto \kappa$) and cantilever damping ($\propto \Gamma$)
we have to study the time evolution of the density matrix $\rho(t)$ with the quantum-optical master equation 
\begin{equation}
\dfrac{\dd \rho}{\dd t} = -\frac{\ii}{\hbar} \, [H,\rho] + \Gamma \mathcal D [\bn,\rho]
                          + \kappa \mathcal D [\an,\rho] \;.
\label{eq:master}
\end{equation}
Note that we work here and in the following at zero temperature,
 such that the dissipative Lindblad terms
\begin{equation}
\mathcal D [L,\rho] = L \rho L^\dagger - \dfrac{1}{2} (L^\dagger L \rho + \rho L^\dagger L)
\label{eq:dissipator}
\end{equation}
contain only bosonic annihilation operators
$L \in \{a, b\}$.

We now express the system parameters in units of $\Omega$,
measure time as $\tau = \Omega t$,
and introduce the two dimensionless parameters~\cite{MHG06, LKM08}
\begin{equation}\label{eq:NewParams}
 P = \dfrac{8 \alpha_L^2 g_0^2}{\Omega^4} \;,
\quad
\sigma = \dfrac{g_0}{\kappa} \;.
\end{equation}
The pump parameter $P$ gives the strength of the laser pumping of the cavity.
The quantum-classical scaling parameter $\sigma=x_\mathrm{zpt}/x_\mathrm{res}=g_0/\kappa$ relates the zero-point fluctuations $x_\mathrm{zpt}=\sqrt{\hbar/(2 m \Omega)}$ of the cantilever (with mass $m$) to the resonance width $x_\mathrm{res}$ of the cavity~\cite{LKM08}.
Note that $x_\mathrm{res}$ is a classical quantity,
characterizing the cavity quality, while $x_\mathrm{zpt}$ is of order $\hbar^{1/2}$ such that $\sigma$ vanishes for $\hbar \to 0$.
Variation of $\sigma$ thus allows us to track how the quantum dynamics of the optomechanical system evolves towards the classical dynamics in the bad-cavity limit $x_\mathrm{zpt} \ll x_\mathrm{res}$, i.e., $\sigma \ll 1$.

For the numerical results we fix the damping parameters $\kappa/\Omega = 1$, $\Gamma/\Omega = 10^{-3}$, 
which are typical values realized in experiments~\cite{AKM13}.
This leaves us with the three parameters $\Delta,  P, \sigma$.

\begin{figure}
\hspace*{\fill}
\includegraphics[width=0.45\linewidth]{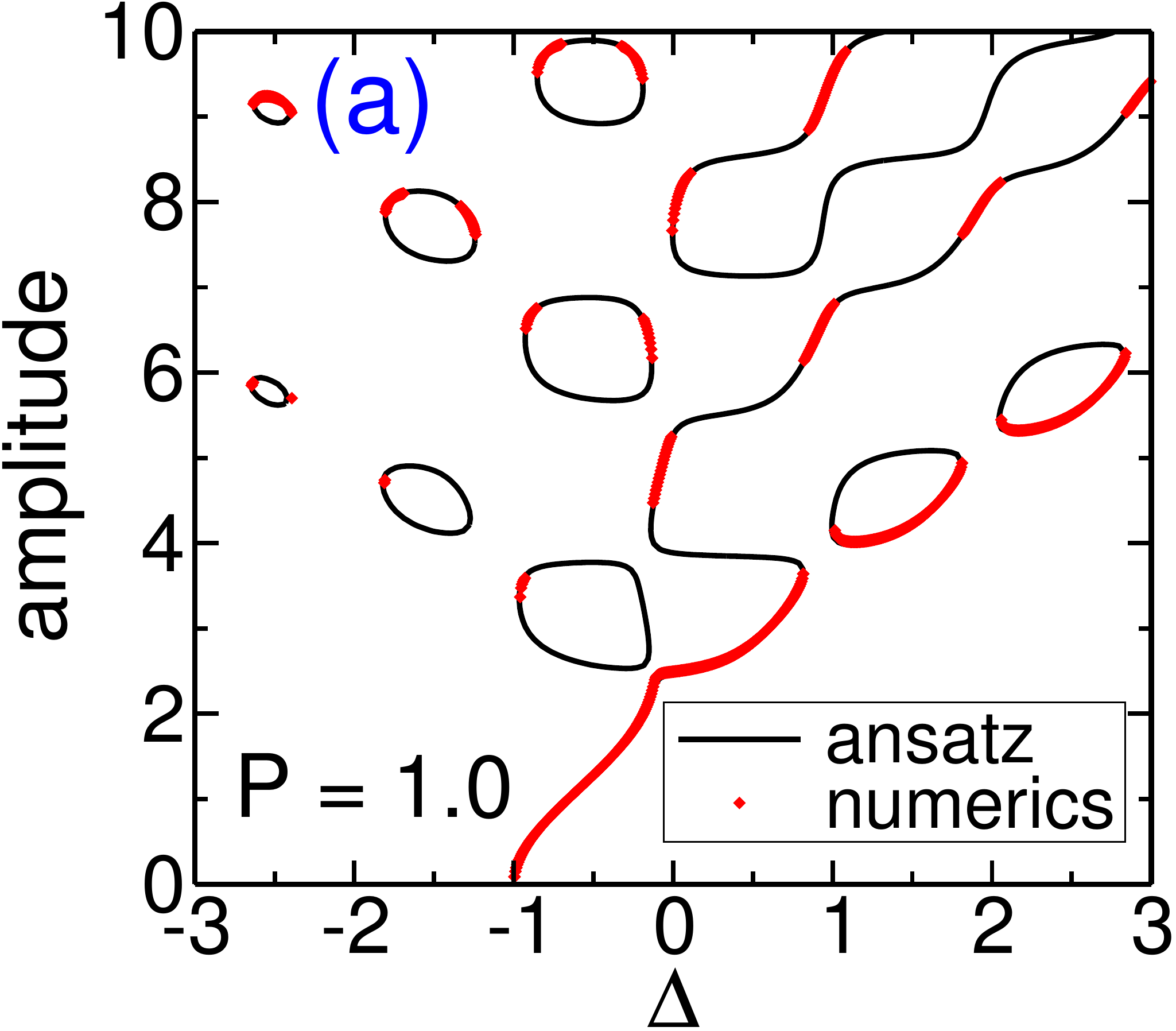}
\hspace*{\fill}
\includegraphics[width=0.45\linewidth]{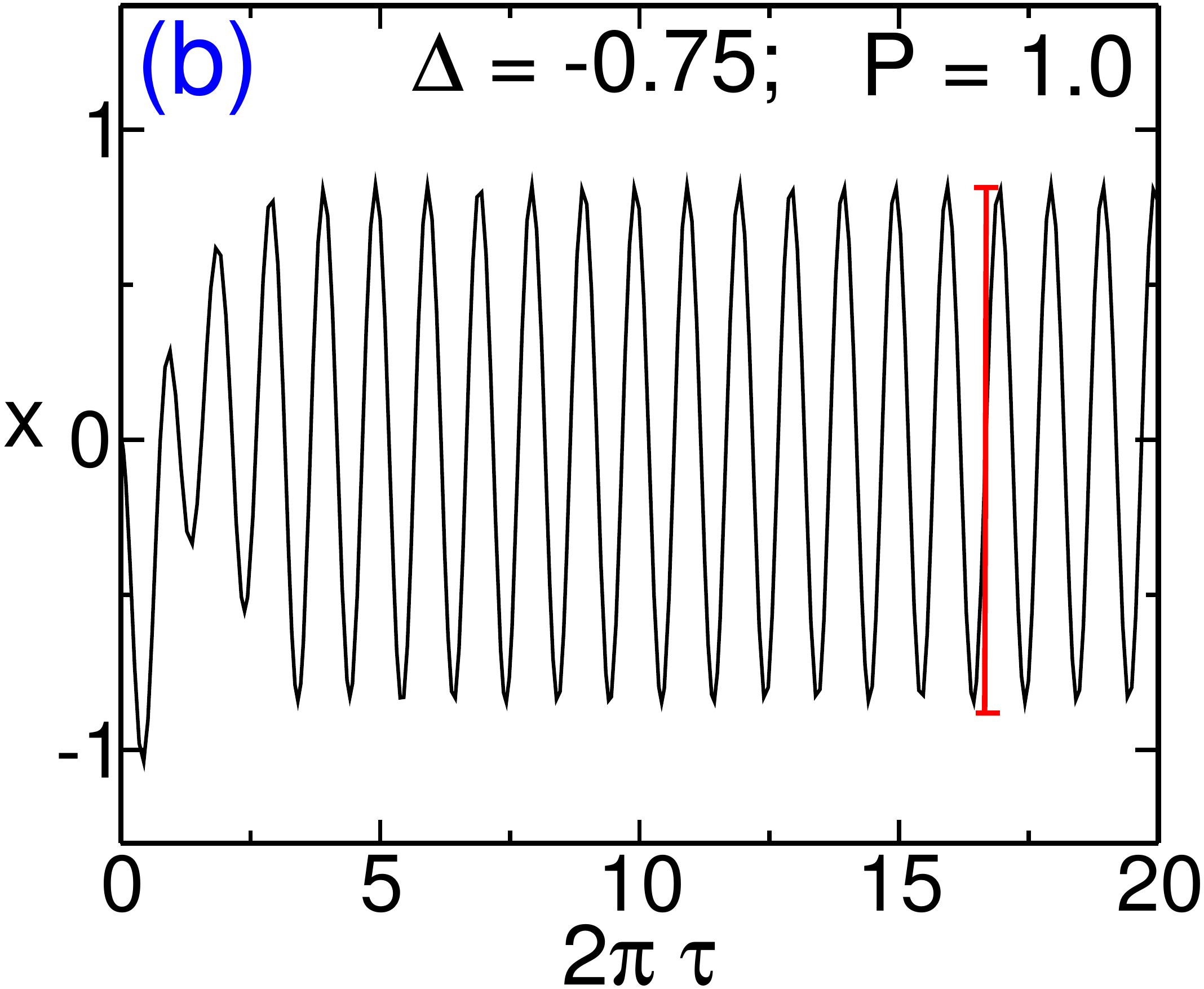}
\hspace*{\fill}
\\
\hspace*{\fill}
\includegraphics[width=0.45\linewidth]{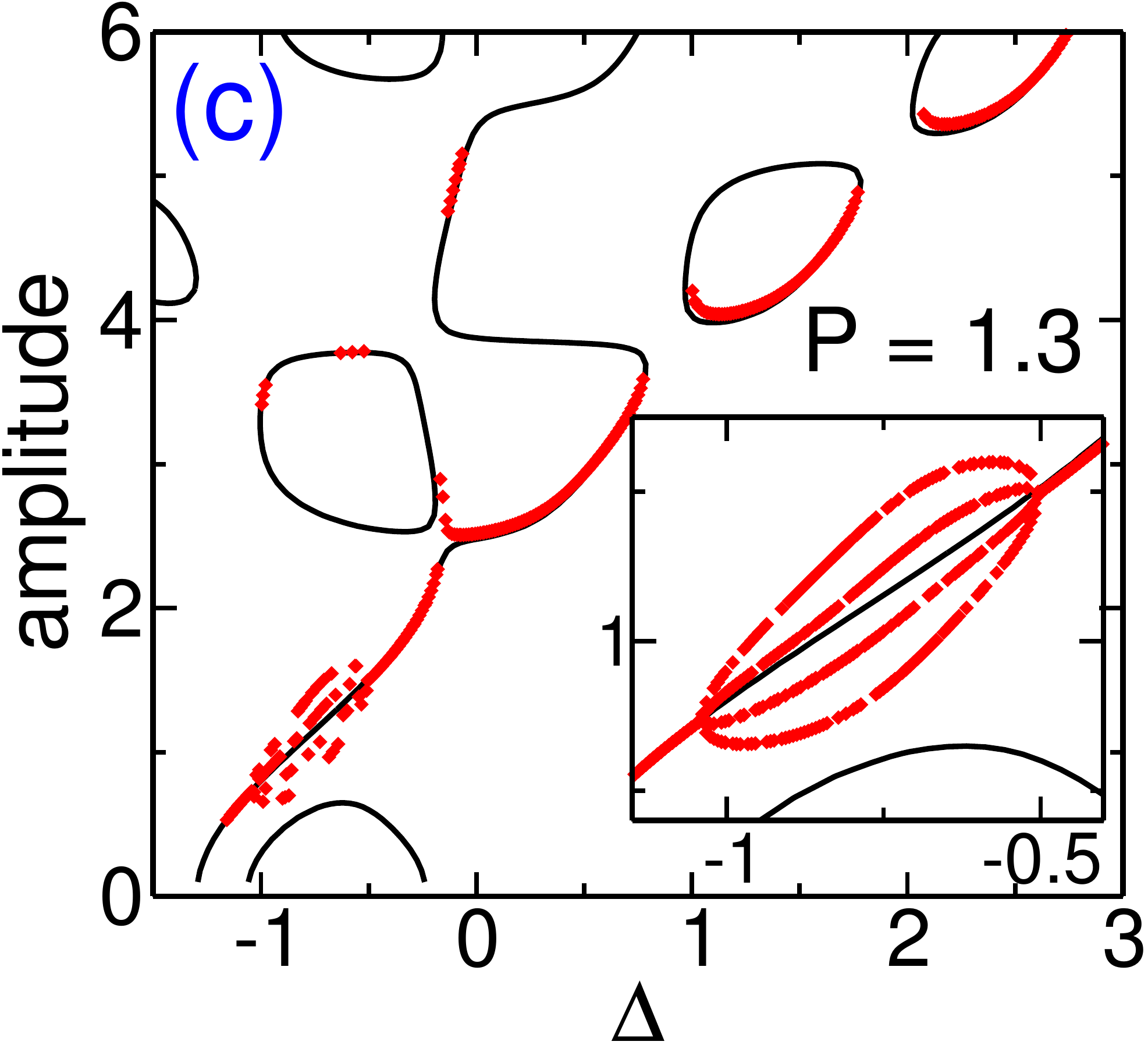}
\hspace*{\fill}
\includegraphics[width=0.45\linewidth]{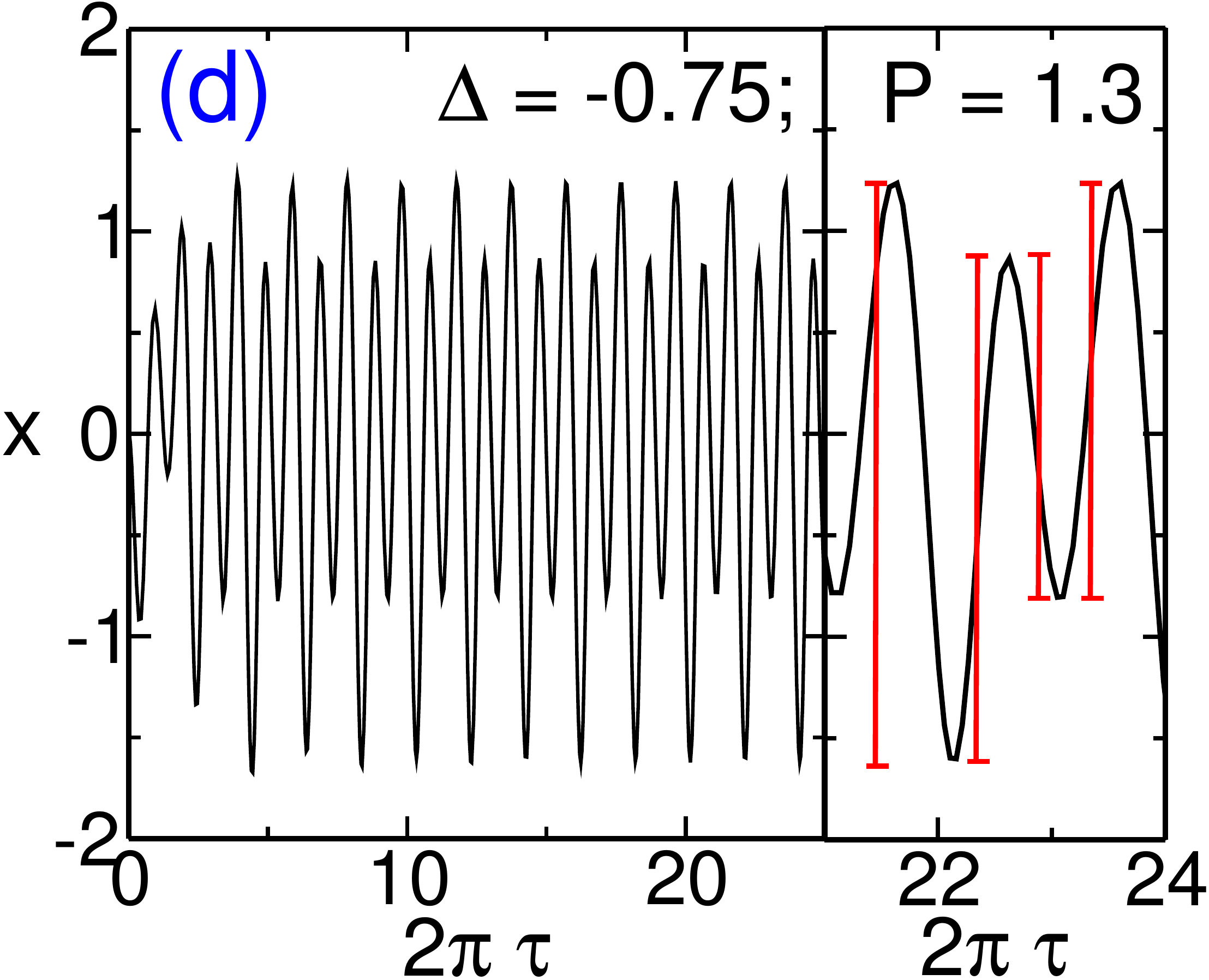}
\hspace*{\fill}
\caption{
(Color online)
(a) and (c): Amplitudes of cantilever oscillation limit cycles given by the sinusoidal ansatz (black line)
and by the full SC equations of motions~\eqref{eq:SC}
(red diamonds).
The inset in (c) displays a zoom into the first PDB around $\Delta=-0.75$.
(b) and (d): Initial dynamics of the cantilever converging to a period-1 resp. period-2 limit cycle.
Here, as in all figures, we give $\Delta, \tau$ in units of $\Omega^{(-1)}$ and use the dimensionless parameters $P$, $\sigma$ from Eq.~\eqref{eq:NewParams}.
}
\label{fig:pdb_01}
\end{figure}

We first establish
the emergence of chaotic motion in the bad-cavity limit $\sigma \ll 1$.
In this limit, the dynamics of the optomechanical systems follows the semi-classical (SC) equations of motion~\cite{LKM08}
\begin{align}
\frac{\dd \alpha}{\dd \tau} & = \phantom{-} \ii \left[ \frac\Delta\Omega  \alpha - (\beta + \beta^*) \alpha
                                              - \frac12 \right]
                                   - \frac{\kappa}{2\Omega} \alpha \;, \\[0.25ex]
\frac{\dd \beta}{\dd \tau} & = -\ii \left[ \frac{P}{2} |\alpha|^2 + \beta \right] - \frac{\Gamma}{2\Omega} \beta 
\label{eq:SC}
\end{align}
for the rescaled cavity and cantilever amplitude
$\alpha = (\Omega/(2\alpha_L)) \langle a \rangle$,
$\beta = (g_0/\Omega) \langle b \rangle$.
For the cantilever we also use the phase space variables $x = 1/\sqrt{2}\, (\beta + \beta^*)$ and $p = -\ii/\sqrt{2}\, (\beta^* - \beta)$.

The SC equations of motion are obtained from the Ehrenfest equations of motion
for the photon $(a^{(\dagger)})$ and phonon $(b^{(\dagger)})$ mode,
together with the SC approximation $\langle (\bd + \bn)\,\an \rangle \approx \langle \bd + \bn\rangle \langle \an\rangle$ in
which all photon-phonon correlations are neglected.
For $\sigma > 0$ the SC equations are an approximation to the full quantum dynamics in Eq. \eqref{eq:master}, but become exact in the limit $\sigma \rightarrow 0$.

The SC equations of motion predict the optical bistability of the optomechanical system,
where self-induced cantilever oscillations arise through a Hopf bifurcation~\cite{MHG06, LKM08}.
The stable attractors of self-induced oscillations are shown in Fig.~\ref{fig:pdb_01}~(a).
The oscillations can be described with a simple sinusoidal ansatz
$x(t) = \bar{x} + A\cos(\Omega t)$
for the cantilever position,
which allows for an analytical solution in terms of a Fourier series~\cite{MHG06,LKM08}.
The predictions of the ansatz agree well with the amplitudes
extracted from the numerical solution of the SC equations (see Fig.~\ref{fig:pdb_01}~(b) for a sample trajectory).

\begin{figure}
\centering
\includegraphics[width=0.75\linewidth]{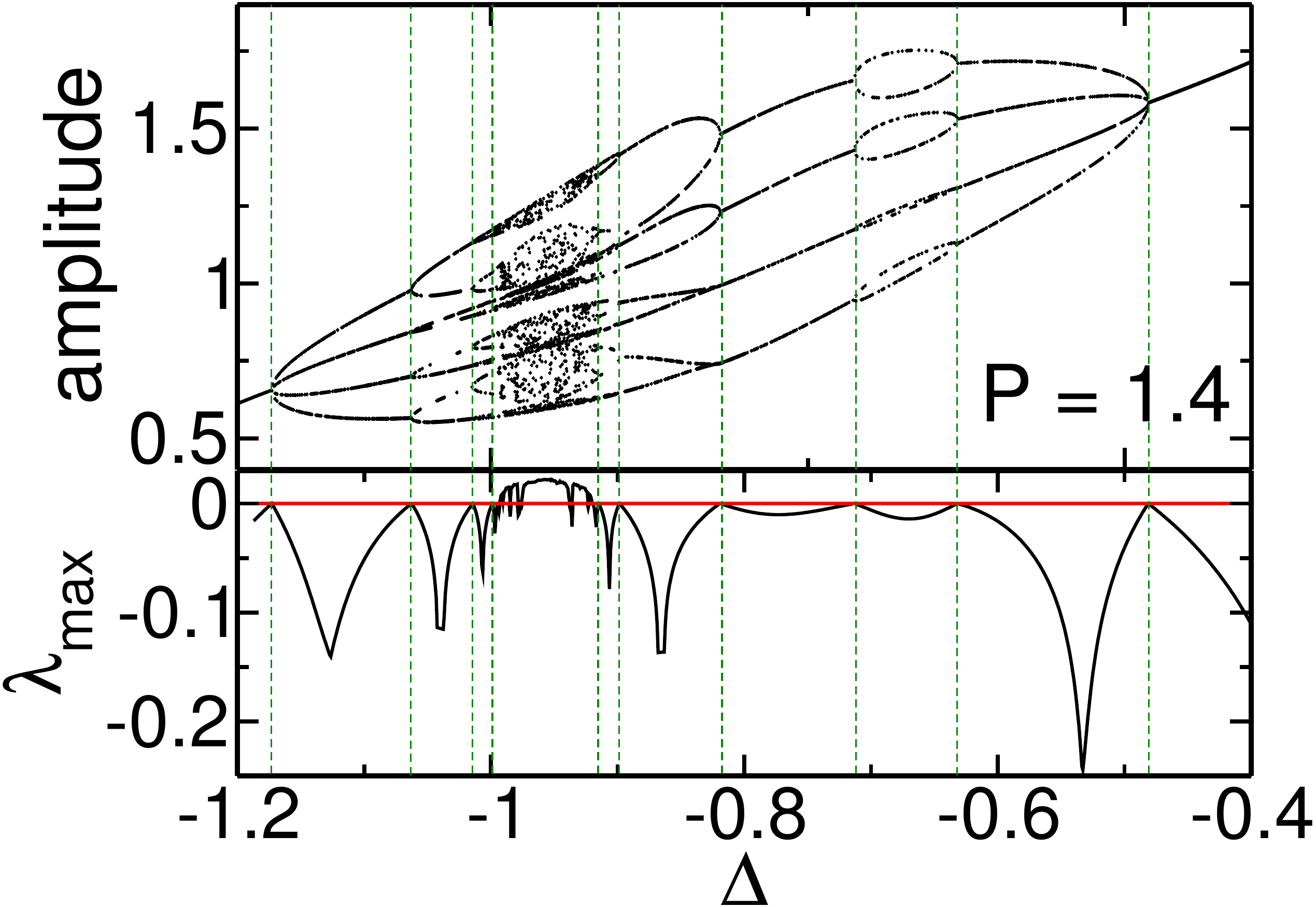}
\caption{(Color online) Bifurcation diagram of the limit cycle amplitude (above) and corresponding maximal Lyapunov exponent $\lambda_\text{max}$ (below) at $ P = 1.4$.
Vertical dashed lines mark PDBs, signaled by $\lambda_\text{max} = 0$.}
\label{fig:pdb_02}
\end{figure}

We now follow the route from regular self-induced cantilever oscillations into the chaotic regime by increasing the pump power $ P$.
For $ P = 1.3$ a period doubling bifurcation (PDB) has taken place,
and a new limit cycle with twice the period of the original simple periodic cycle appears for negative detuning and small amplitude, as shown in Fig.~\ref{fig:pdb_01}~(c).
A sample trajectory located on the period-2 limit cycle is shown in Fig.~\ref{fig:pdb_01}~(d).
The single-frequency ansatz fails trivially predicting the PDB, the 
four possible ``amplitudes'' of the period-2 cycle are extracted
from the numerical solution of the SC equations.

Increasing the driving further leads to additional PDBs
and the appearance of period-$n$ limit cycles (not shown here).
Eventually, for $ P=1.4$, chaotic motion emerges
as shown in Fig. \ref{fig:pdb_02}.
We distinguish chaotic and regular trajectories through the maximal Lyapunov exponent (LE), which we calculate with the ``standard'' method from \cite{BGGSa80, BGGSb80}.
The LE vanishes at every PDB, and separates regular motion with a negative LE
from chaotic motion with a positive LE.
As the LE in Fig. \ref{fig:pdb_02} shows, the chaotic region is bounded and contained within a small window $\Delta \in [-1.0,-0.91]$.

\begin{figure}
\centering
\includegraphics[width=0.8\linewidth]{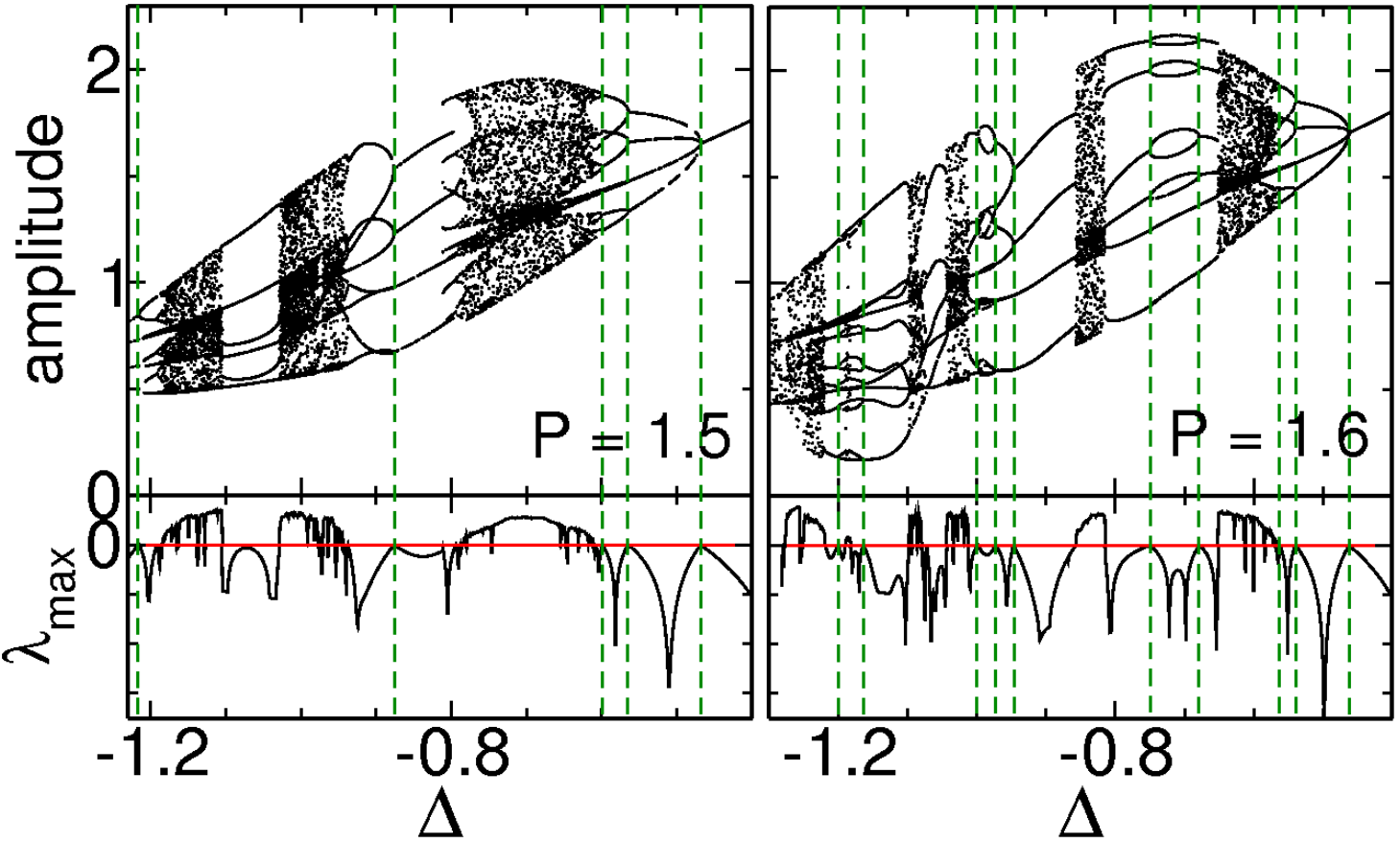}
\caption{(Color online) Bifurcation diagrams of the limit cycle amplitude and maximal Lyapunov exponents $\lambda_\text{max}$ for larger pump power $ P=1.5, 1.6$.}
\label{fig:pdb_03}
\end{figure}

The bifurcation diagrams get more complex with increasing $P$ (Fig. \ref{fig:pdb_03}).
The chaotic regions do not only expand, but they also split and form a fairly complex intertwined sequence of windows of
regular and chaotic dynamics.
Notably, the appearance of regular or chaotic motion is very susceptible to the value of $\Delta$.
Changing the laser-cavity detuning one can easily tune the optomechanical system in and out of chaos.

\begin{figure}[b]
\hspace*{\fill}
\includegraphics[width=0.7\linewidth]{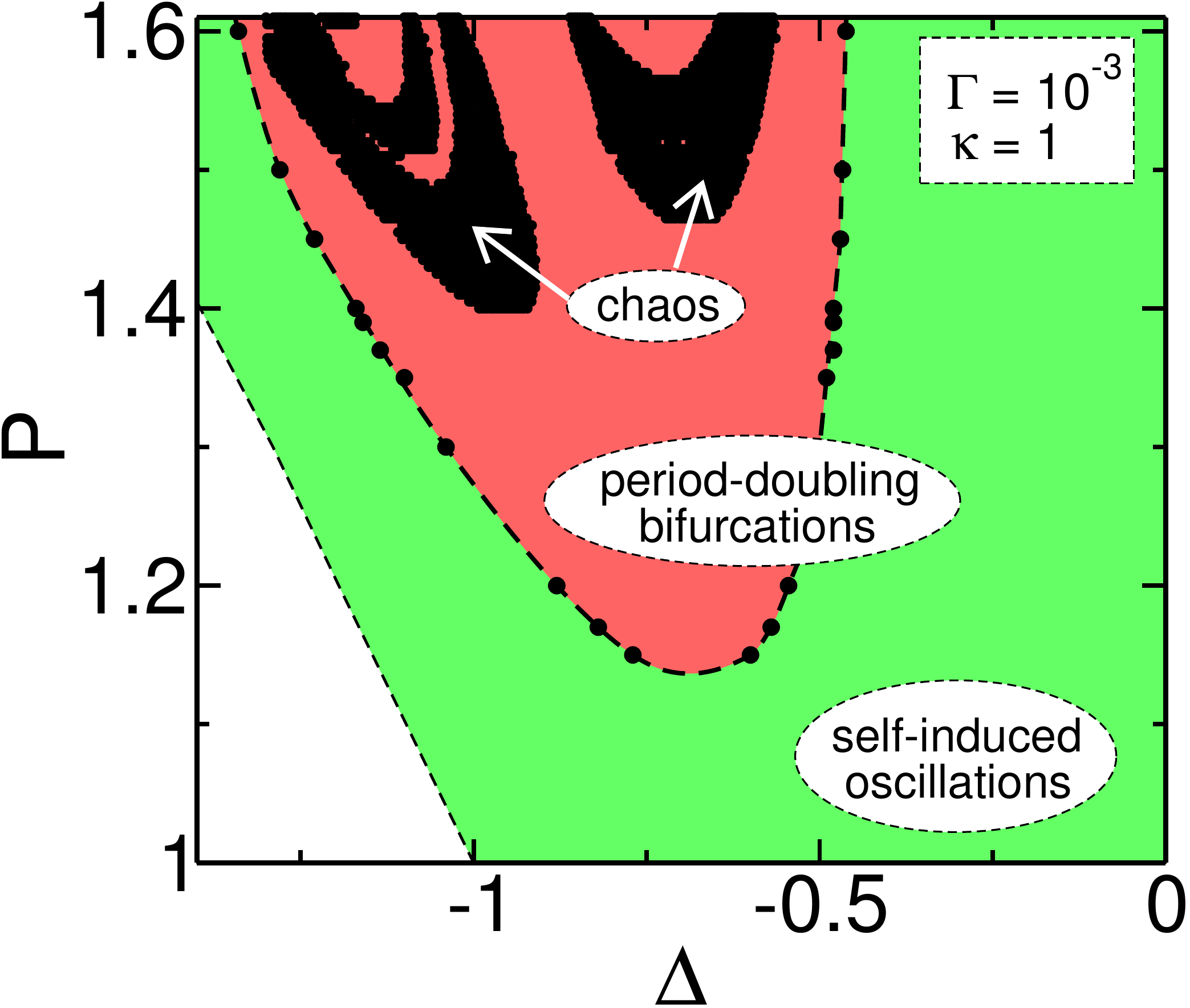}
\hspace*{\fill}
\caption{(Color online) Schematic picture of the regular and chaotic regimes of the optomechanical system in the bad-cavity limit $\sigma \ll 1$, plotted in the $\Delta$-$ P$ plane. 
Here, as everywhere, $\kappa=1$ and $\Gamma=10^{-3}$.
Dots represent numerical data extracted from bifurcation diagrams (such as Figs.~\ref{fig:pdb_02},~\ref{fig:pdb_03}),
dashed lines interpolate between the numerical data.}
\label{fig:Sketch}
\end{figure}

The appearance of chaos 
 is summarized in Fig.~\ref{fig:Sketch}.
If we follow the route to chaos by increasing the pump power $P$, the first PDBs occur in a parabolic region for $P \gtrsim 1.15$ 
 before chaotic motion sets in for slightly larger $P \gtrsim 1.4$.
The chaotic regime does not form a simple convex part of the parameter plane, but has a complex structure characterized by interjacent regions of regular motion.
Importantly, PDBs and chaos appear at parameter values $-1.5 \le \Delta$ and $P \le 1.6$  accessible to experiments.

\begin{figure}
\centering
\includegraphics[width=0.7\linewidth]{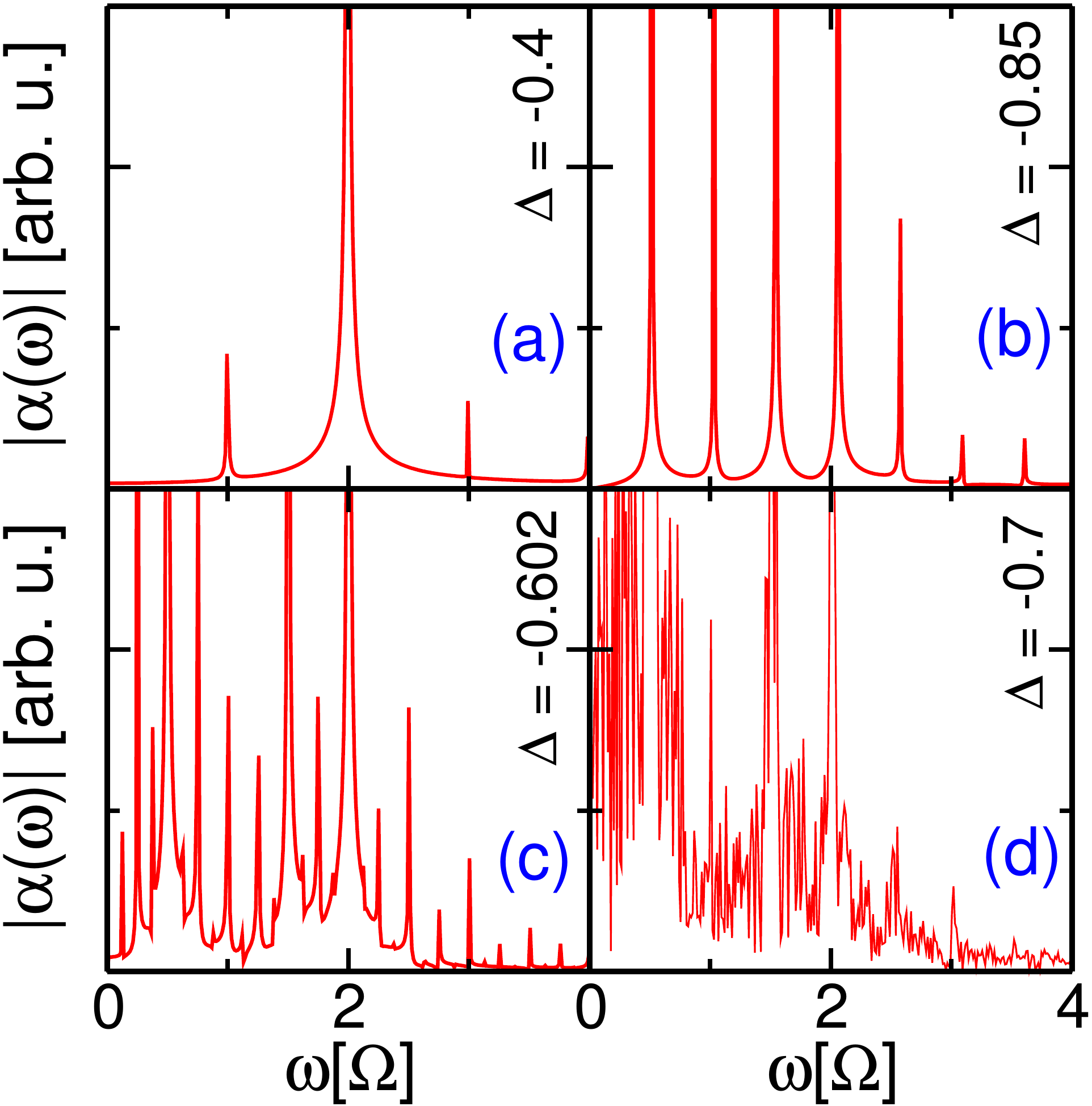}
\caption{(Color online) Fourier spectrum of the classical photon field amplitude $\alpha$ for $ P = 1.5$.
The figures show spectra corresponding to cantilever dynamics on a simple periodic limit cycle (a), a period-2 limit cycle (b),
a period-4 limit cycle (c), and a chaotic limit cycle (d).}
\label{fig:spec_alpha}
\end{figure}

Experimental evidence for chaotic motion can be obtained from the cavity intensity spectrum, as shown in Fig. \ref{fig:spec_alpha}.
For period-1 oscillations, with $\Delta=-0.4$ to the right of the chaotic window in Fig.~\ref{fig:pdb_02}, peaks in the spectrum occur only at multiples of the cantilever frequency $\Omega$ (panel (a)).
Moving further into the negative detuning regime ($\Delta < -0.4$),
additional peaks occur between the peaks of the preceding spectrum with each PDB, at multiples of $\Omega/2$ ($\Omega/4$) after the first (second) PDB in panel (b) (panel (c)), until the chaotic regime is reached and the spectrum becomes continuous (Fig.~\ref{fig:spec_alpha}~(d)).

Note that we focus on negative detuning $\Delta < 0$,
where chaos appears already at moderate pump power $P$.
Chaotic motion exists also for positive detuning $\Delta > 0$,
but then requires much larger $P$ such that it will be harder to access experimentally.

We now turn to the quantum dynamics of the optomechanical system,
which we compute with the quantum state diffusion method (QSD)~\cite{GP92, D88}
using the implementation of Ref.~\cite{SB97}.
In QSD the density matrix $\rho(t)$ 
is represented by an ensemble of individual quantum trajectories,
which evolve according to a stochastic differential equation 
that replaces the master equation~\eqref{eq:master}.
One advantage of QSD over, e.g., the quantum jump method \cite{MCD93} 
is the dynamical localization of quantum trajectories on classical orbits~\cite{RG96, SBP95, SP98}.
Therefore, the emergence of classical from quantum dynamics in the bad-cavity limit $\sigma \ll 1$ can be observed directly through comparison of individual classical and quantum trajectories.

\begin{figure}
\centering
\includegraphics[width=0.7\linewidth]{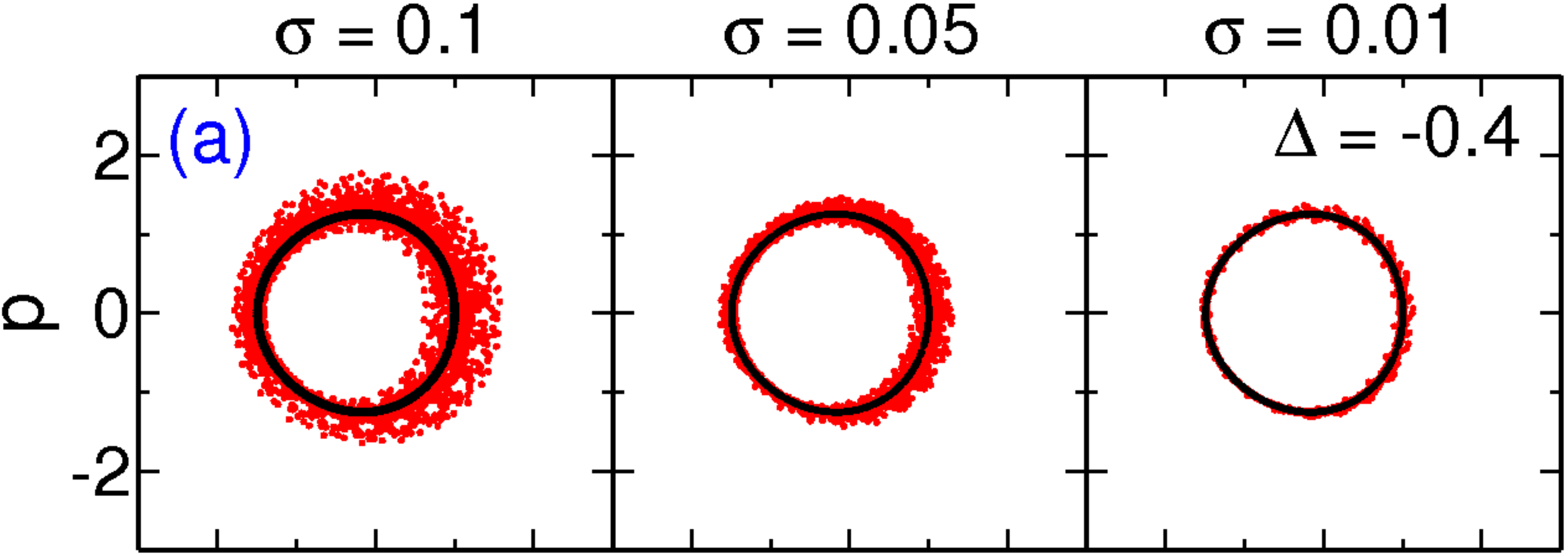} \\
\includegraphics[width=0.7\linewidth]{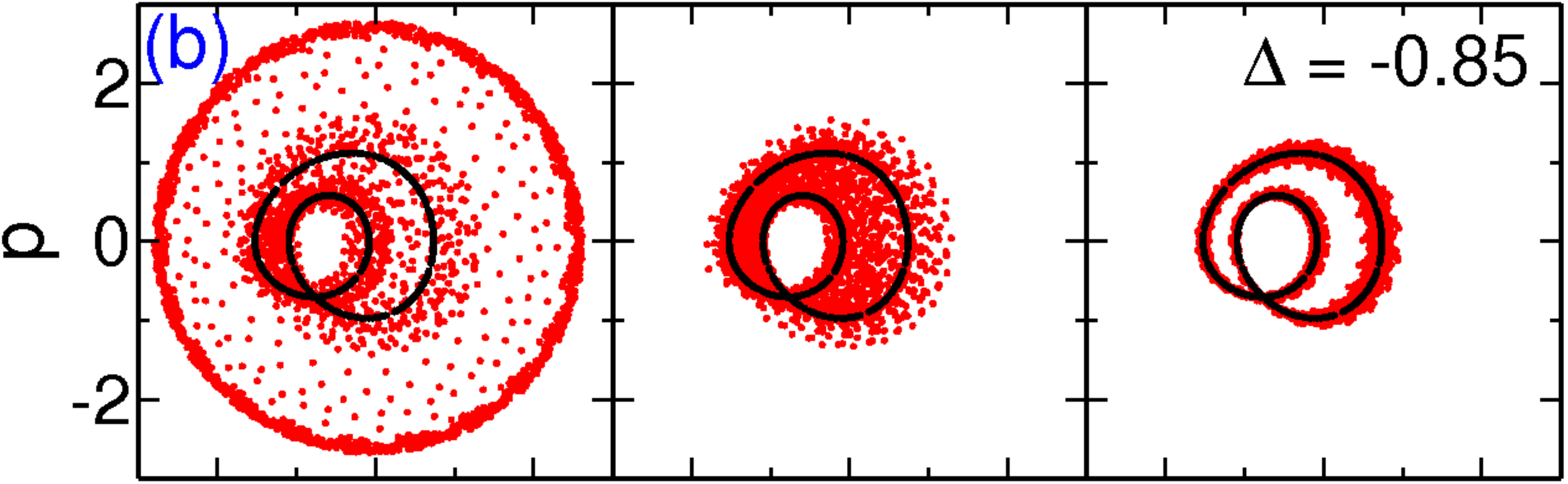} \\
\includegraphics[width=0.7\linewidth]{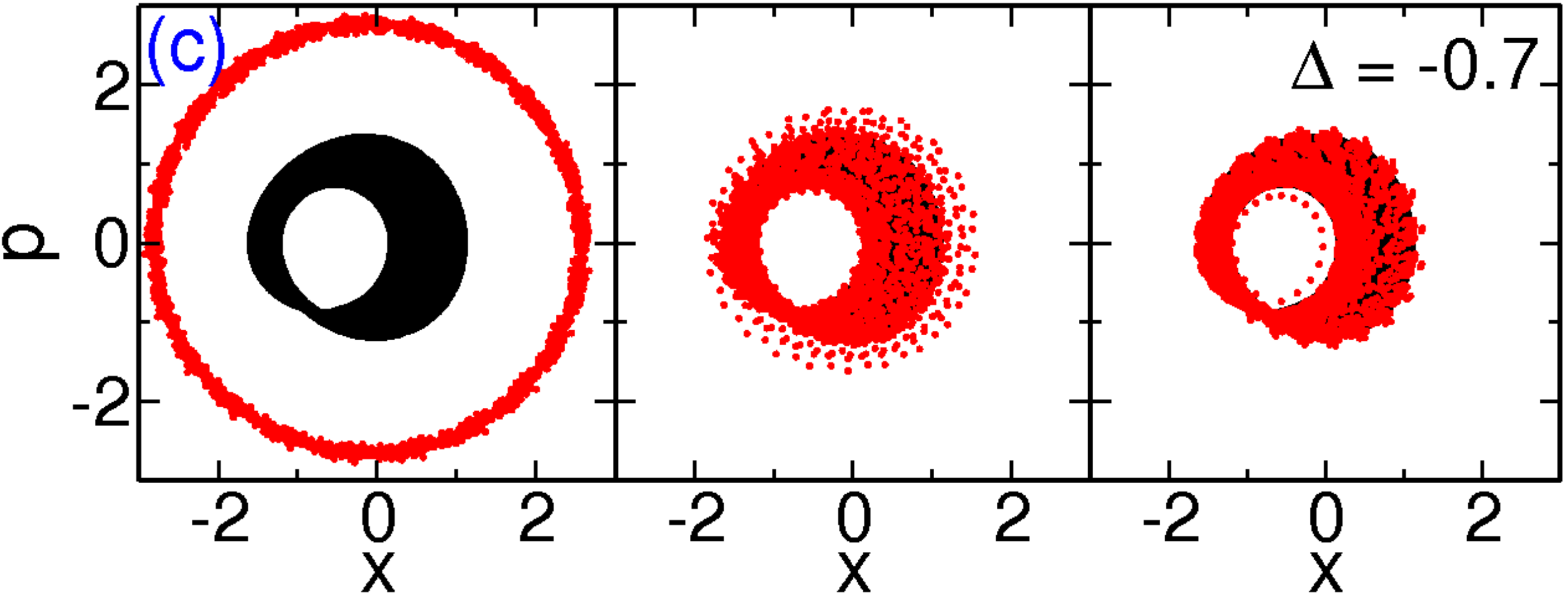}
\caption{(Color online) Stroboscopic $(x,p)$-phase space plot of a single quantum trajectory (red points) 
for $P=1.5$, initial conditions $(x_0,p_0)=(0,0)$,
and decreasing
quantum-classical scaling parameter $\sigma$, approaching a simple periodic limit cycle (a), a period-2 limit cycle (b) and a chaotic limit cycle (c) (black curves) in the limit $\sigma \to 0$.
We show points for $\tau/(2\pi) \ge 20$ only to omit the earliest transient dynamics.
}
\label{fig:qm_singleMC}
\end{figure}

Typical quantum trajectories 
are shown in Fig.~\ref{fig:qm_singleMC}.
We  observe how with decreasing $\sigma$ the trajectories localize on the classical limit cycles. 
More complex classical orbits require smaller values of $\sigma$ for localization.
The localization properties change in the quantum regime ($\sigma = 0.1$, panels (b), (c)).
Now the quantum trajectory leaves the classical limit cycle
and localizes on a different simple periodic orbit.
The crossover to the new orbit happens more quickly for the chaotic limit cycle (panel (c)).

We note that classical periodic orbits exist in the vicinity of the new quantum orbit, but the value $\sigma=0.1$ is already too large for a definite assignment of classical to quantum trajectories.
Further work is necessary to decide whether the quantum trajectory localizes on different classical orbits as $\sigma$ is increased, or whether the new quantum orbit appears because the quantum-classical correspondence~\cite{BRA11} breaks down entirely.

\begin{figure}
\centering
\includegraphics[width=0.45\linewidth]{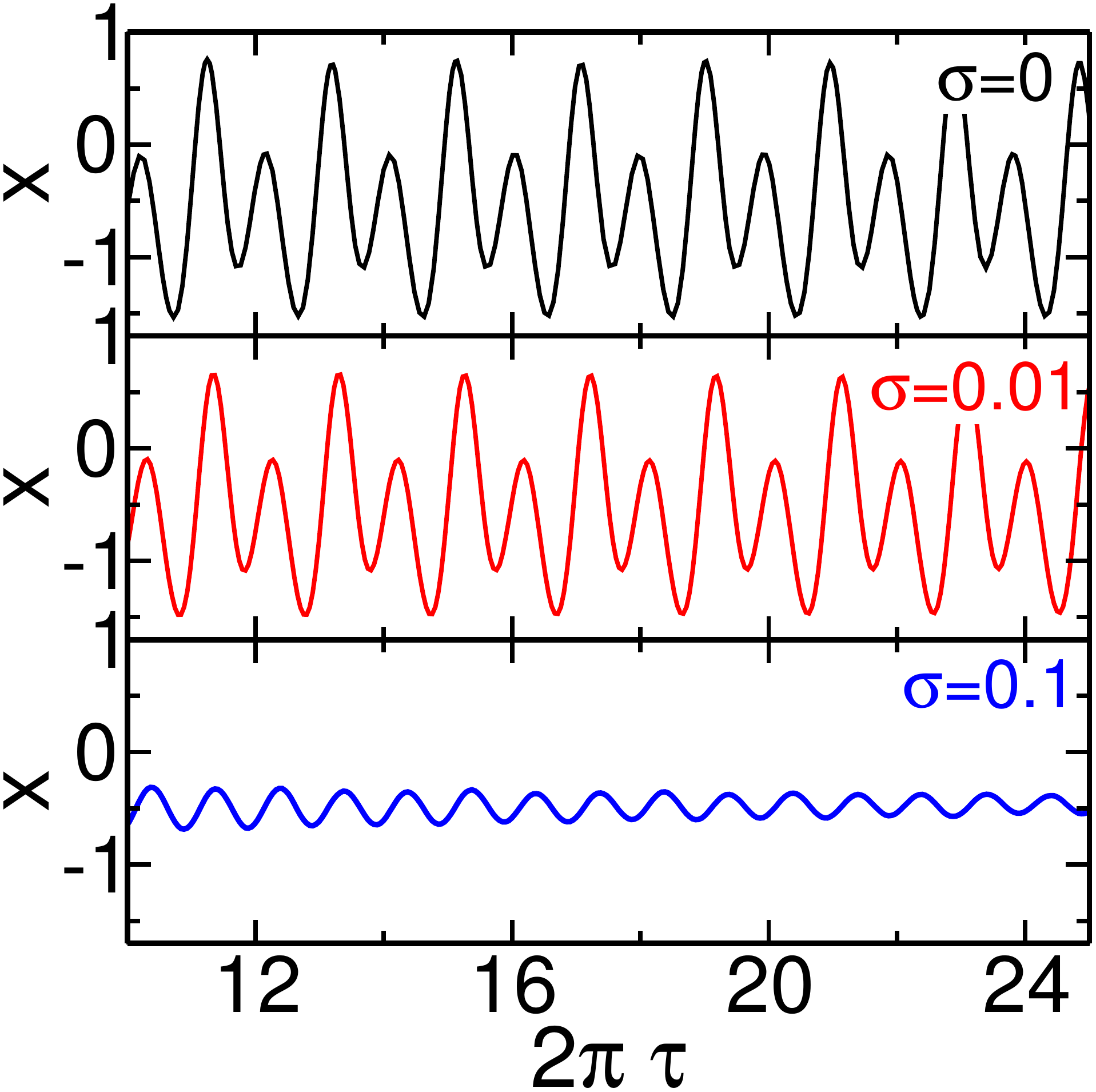}
\hfill
\includegraphics[width=0.45\linewidth]{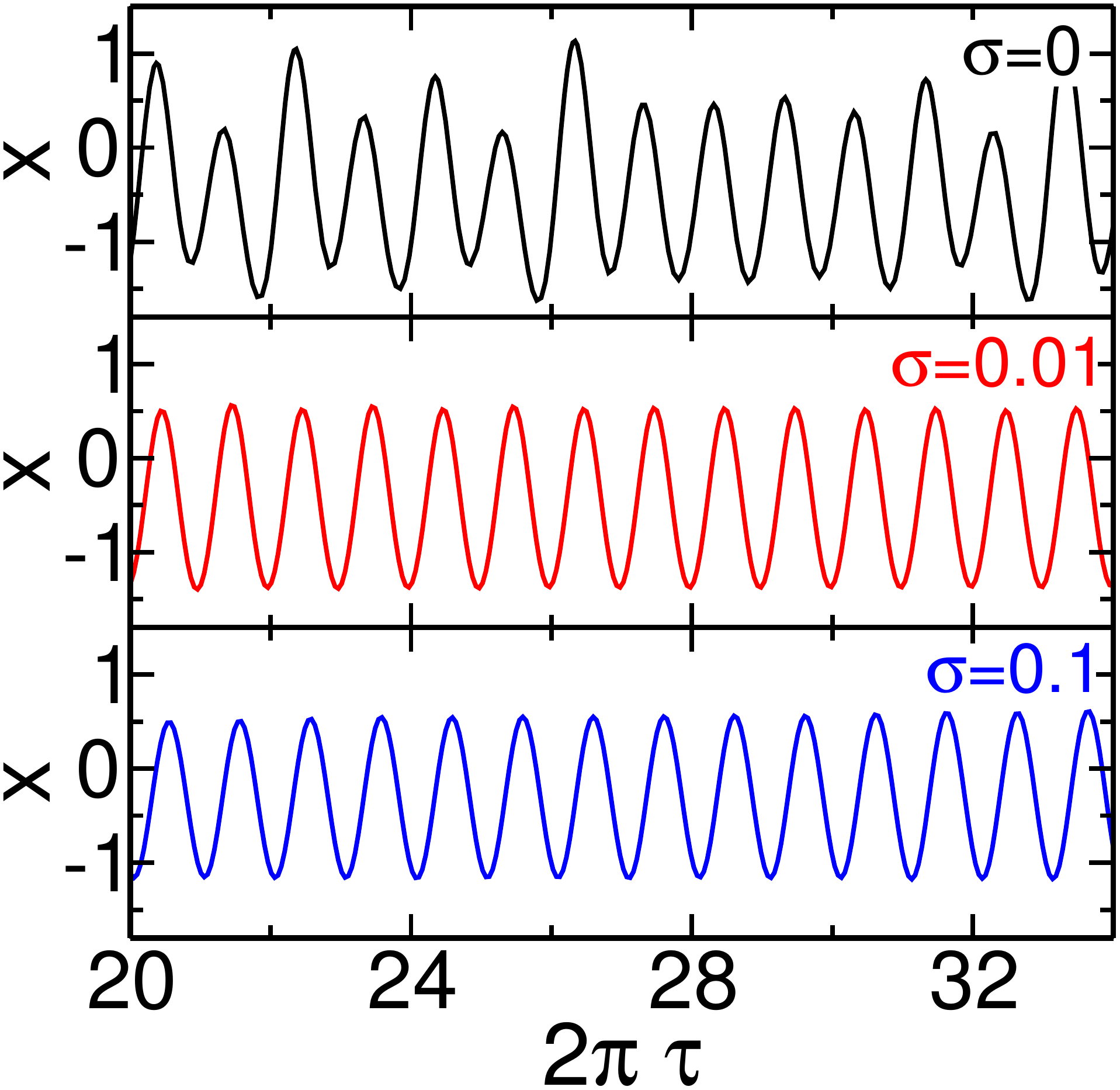}
\caption{(Color online) Quantum dynamics of the cantilever from the ensemble average of 5000 quantum trajectories ($\sigma > 0$)
in comparison to the SC dynamics ($\sigma = 0$), all for $ P=1.5$.
The figures depict the case of a classical period-2 orbit ($\Delta = -0.85$, left) and a chaotic orbit ($\Delta = -0.7$, right).
}
\label{fig:qm_ensemble}
\end{figure}

The behavior of individual quantum trajectories
provides only a qualitative idea of the quantum dynamics.
Experimentally accessible quantities are obtained from 
the ensemble average over all trajectories.
Fig. \ref{fig:qm_ensemble} shows the resulting cantilever position $x(t)$
in comparison to the SC trajectories after the initial transient dynamics has faded out.
As the system evolves out of the SC limit into the quantum regime one observes that the classical (period-2 or chaotic) motion is replaced by simple periodic cantilever oscillations.
As anticipated from the localization properties of the individual quantum trajectories the quantum dynamics favors simple periodic motion.

The two curves for $\sigma=0.01$ in Fig.~\ref{fig:qm_ensemble} indicate
that the position of the crossover from classical to period-1 motion depends on the complexity of the classical limit cycle. For chaotic orbits it takes place closer to the SC limit, 
such that the $\sigma=0.01$ curve still agrees with the classical dynamics for a period-2 orbit (left panels),
but already  shows the simple periodic oscillations of the quantum regime for a classically chaotic orbit (right panels).

To conclude, we here analyze the route to chaos in the optomechanical system
and track the appearance of PDBs in the cantilever oscillations as the first step to mixed regular-chaotic dynamics.
From comparison of the SC and quantum dynamics we find
evidence that the quantum dynamics is protected against chaotic motion.
The central observation is that quantum mechanics seems to counteract the 
classical route to chaos in the optomechanical system and stabilizes simple periodic orbits. 
%This behavior can be traced back to the different localization properties of individual quantum trajectories.

The present analysis is carried out at zero temperature, which is justified 
because in the experiments on chaos in the optomechanical system~\cite{CRYKV05, CCV07} the temperature is small on the relevant energy scale set by the self-induced oscillations.
Because the temperature of the optomechanical system can be controlled effectively~\cite{T_etal11, C_etal11}
an analysis of the influence of small finite temperatures on the correspondence between classical and quantum dynamics~\cite{KRSL07,KLRS08} in the chaotic regime should be rewarding for future theoretical and experimental studies.

Besides being of interest in itself,
the existence of chaos in the optomechanical system could be relevant for ultra-precision measurements or fundamental tests on the physical conditions for classical dynamics.
Because the mixed regular-chaotic dynamics we depicted
is susceptible to small variations of the systems parameters, e.g., the cantilever mass
or the laser-cavity detuning, such variations can be detected through drastic changes in the cantilever dynamics.
On the other hand, the fact that the quantum dynamics favors simple periodic over multi-periodic or irregular chaotic  motion may help to explain why optomechanical systems can be used in a controlled way even deep in the quantum regime.

First experimental results on the observation of PDBs and chaotic motion in
the intensity spectrum of an optomechanical system were reported in~\cite{CRYKV05, CCV07}.
In light of our results we suggest to continue experimental studies 
in this direction, systematically tracing out the boundaries of the regular and chaotic regimes in comparison to the theoretical predictions garnered from the SC equations of motion.
In particular, one should try to locate the crossover from classically multi-periodic or chaotic motion to the simple periodic quantum dynamics by changing the quantum-classical scaling parameter $\sigma$, e.g., through variation of the
 cantilever mass.

\begin{acknowledgments}
{\it Acknowledgments.}
The authors wish to thank B. Bruhn and F. Marquardt for helpful discussion.
This work was supported by Deutsche Forschungsgemeinschaft via SFB 652 (B5) and 
Priority Programme 1648 "Software for Exascale Computing".
\end{acknowledgments}

\end{document}